\documentclass[english,twocolumn]{revtex4}
\usepackage[T1]{fontenc}
\usepackage[latin9]{inputenc}
\usepackage{graphicx}
\usepackage{babel}

\usepackage{subfig}
\usepackage[subfigure]{tocloft} 

\usepackage{fancyvrb}
 \usepackage{mathrsfs}
\usepackage{amsmath}
\usepackage{amssymb}
\usepackage{wasysym}
\usepackage{graphicx}
\usepackage{esint}
\usepackage{dcolumn}

\begin{document}

\title{Gravitational Lensing as a Mechanism For Effective Cloaking}
\author{Benjamin K. Tippett}
\email{bktippett@mun.ca}

\affiliation{Department of Mathematics and Statistics  \\ Memorial University of Newfoundland\\ St. John's, NL, A1C 5S7 \\ Canada}

\begin{abstract}
In light of the surge in popularity of electromagnetic cloaking devices,
we consider whether it is possible to use general relativity to \emph{cloak}
a volume of spacetime through gravitational lensing. A metric for such a spacetime geometry is presented,
and its geometric and physical implications are explained.
\end{abstract}
\maketitle

\section{Introduction}

In general relativity, there is a tradition of engineering spacetime
geometries with exotic attributes previously seen only in science
fiction. Tipler \cite{tipler} and Morris \cite{morris} have introduced
time machines; and Alcubierre \cite{alcubierre} introduced a warp
drive. 

In science fiction, one popular conceit is the idea of a cloaking
device: a mechanism through which a spaceship could be made undetectable.
The revelation that curved spaces can be matched to the electromagnetic
properties of a medium has sparked a recent interest in optical cloaking
\cite{ahn,Crudo,leonhard,leonhard2,leonhardXX,pendry,Schurig}. We
seek to construct a spacetime geometry which cloaks an interior region
from null geodesics.

\subsection{Electromagnetic Cloaking}

A lot of interest has recently been awakened in researching cloaking
devices. Indeed, since it is now possible to engineer metamaterials
with desired exotic electromagnetic properties, a model of a cylindrically
symmetric invisibility cloak has even been assembled by Schurig \emph{et
al.} \cite{Schurig}.

\begin{figure}
\centering \protect\includegraphics[trim=10mm 40mm 130mm 30mm,clip,width=0.45\textwidth]{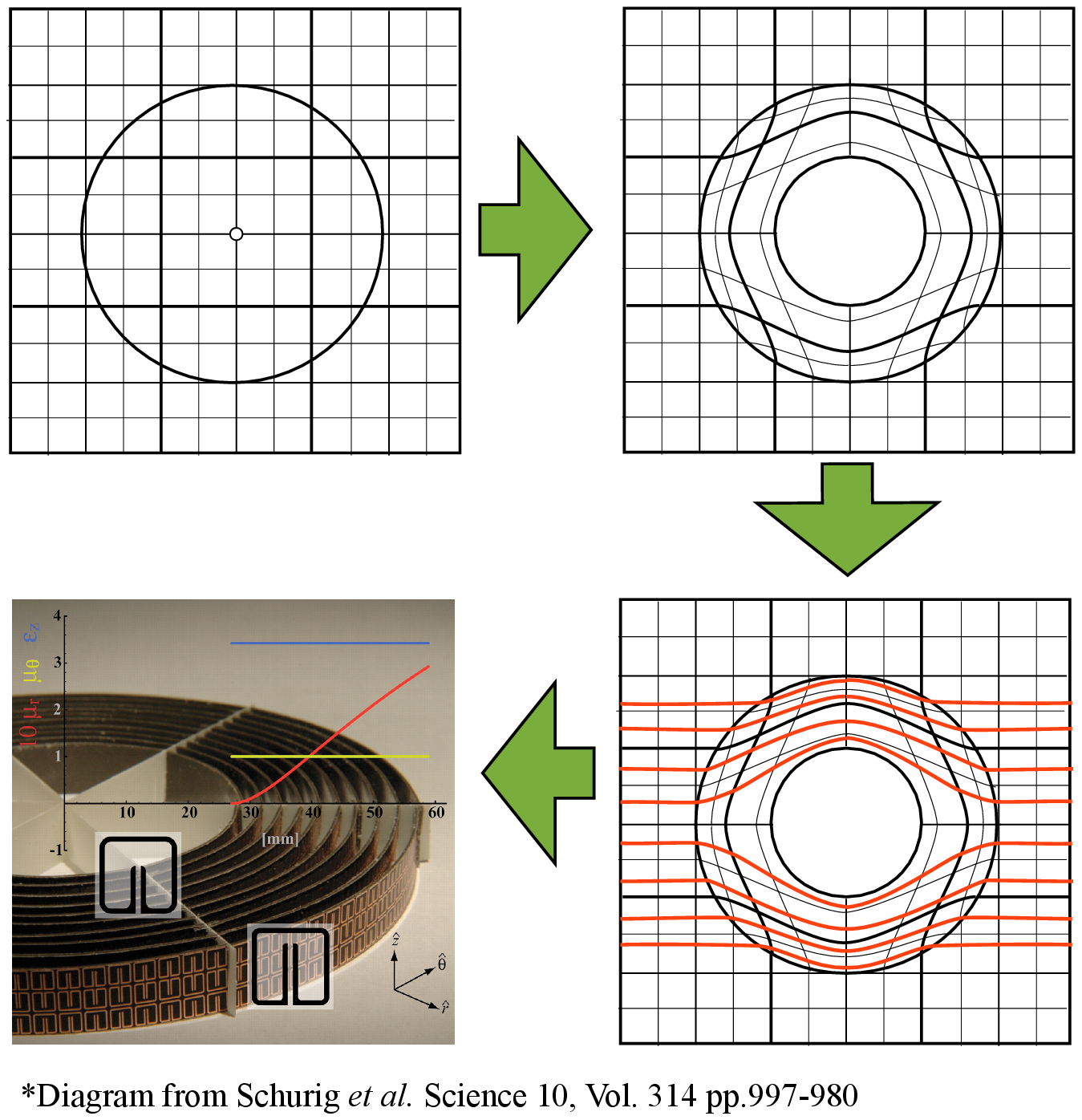}
\caption{The procedure for designing an electromagnetic cloaking device. \label{fig:StolenGraph}}
\end{figure}

These recent results have come about as a result of a technique known
transformation optics which allows us to view refraction equivalently
in terms of geodesics on an \emph{virtual curved electromagnetic space},
or in terms of a varying permittivity and permeability in a material
\cite{pendry,leonhard,leonhard2,leonhardXX}. Thus, engineering
systems with exotic electromagnetic properties becomes a matter of
searching for desirable coordinate transformations of flat space \cite{ahn}. For example, consider the
procedure for designing Schurig \emph{et al}'s cloaking device (see Fig.\eqref{fig:StolenGraph}).
We begin with a flat 2-dimensional space, and we draw a circle. We
perform a coordinate transformation which expands the point $r=0$
into a circle of radius $r_{c}$. The {}``straight line'' geodesics
of the old coordinate system will now follow curved paths which circumnavigate
the circle at $r_{c}$. We then use the transformed metric to determine the permittivity
and permeability tensors required for light rays to follow these curves.
The result is an object which electromagnetically cloaks objects inside of the $r_{c}$ circle from the exterior.

This language of geodesics in curved space has opened a dialogue between
transformation optics and general relativity \cite{leonhard}. Crudo
and O'Brien \cite{Crudo} have even generalized the procedure to consider
curved spacetime: given a desired set of geodesics, they determine
the metric of the curved spacetime required to generate the geodesics,
and then the index of refraction required.

The work done on effective metrics in \emph{virtual} spacetime has
inspired us to consider whether cloaking can be achieved as a result
of the curvature of \emph{physical} spacetime.

\subsection{Effective Cloaking in Gravity}

An optical cloaking system must satisfy two criteria: firstly, an external congruence of
light rays which enter the system must exit the system undistorted;
secondly, these light rays are prevented from penetrating an internal
volume. If we generalize these criteria to general relativity, the
light rays become null geodesics, and consequently the interior volume 
 is causally isolated from the exterior spacetime. 

In lieu of causally isolating the interior volume, let us broaden
our definition for cloaking. If a spacetime geometry contains a region
through which a congruence of null geodesics can pass undistorted,
and if the parameters defining the system can be tuned so that an extended
object placed within said region will appear arbitrarily small from
the outside; we will refer to the region as \emph{effectively cloaked.}
Thus, we could use an effective cloaking geometry to make an object
the size of the planet Jupiter appear from the outside to be the size of a pea. In some ways, the mechanism will work in a way opposite of a microscope.

\section{Geomteric Cloaking}

\subsection{Interior and Exterior Metric}
Our cloaking geometries are a two parameter family of geometries constructed
by joining a flat \emph{exterior} spacetime to a curved \emph{interior} spacetime along a spherical hypersurface using the Israel junction conditions \cite{poisson}. 

Let the flat exterior have line element:
\[
ds^{2}=-dt^{2}+dr^{2}+r^{2}d\theta^{2}+r^{2}sin^{2}\theta d\phi^{2}\;,
\]
where the junction hypersurface is the sphere of constant coordinate radius $r=R$.

Let the interior spacetime have line element:
\[
ds^{2}=-(\frac{\tilde{r}}{SR})^{2-2S}dt^{2}+d\tilde{r}^{2}+\frac{1}{S^{2}}\tilde{r}^{2}d\theta^{2}+\frac{1}{S^{2}}\tilde{r}^{2}sin^{2}\theta d\phi^{2},
\]
where $S>1$ is a constant, and the junction hypersurface is a sphere of constant coordinate radius $\tilde{r}=SR$.

  Since the metric
is not smooth across this junction, there must be a shell of stress
energy confined to the spherical junction hypersurface.

\subsection{Geodesic Congruences}
\begin{figure} \centering
             \subfloat[Very weak effective cloaking $S=1.1$]{\label{fig:RTclass} \includegraphics[trim=10mm 10mm 2mm 10mm,clip,width=0.21\textwidth]{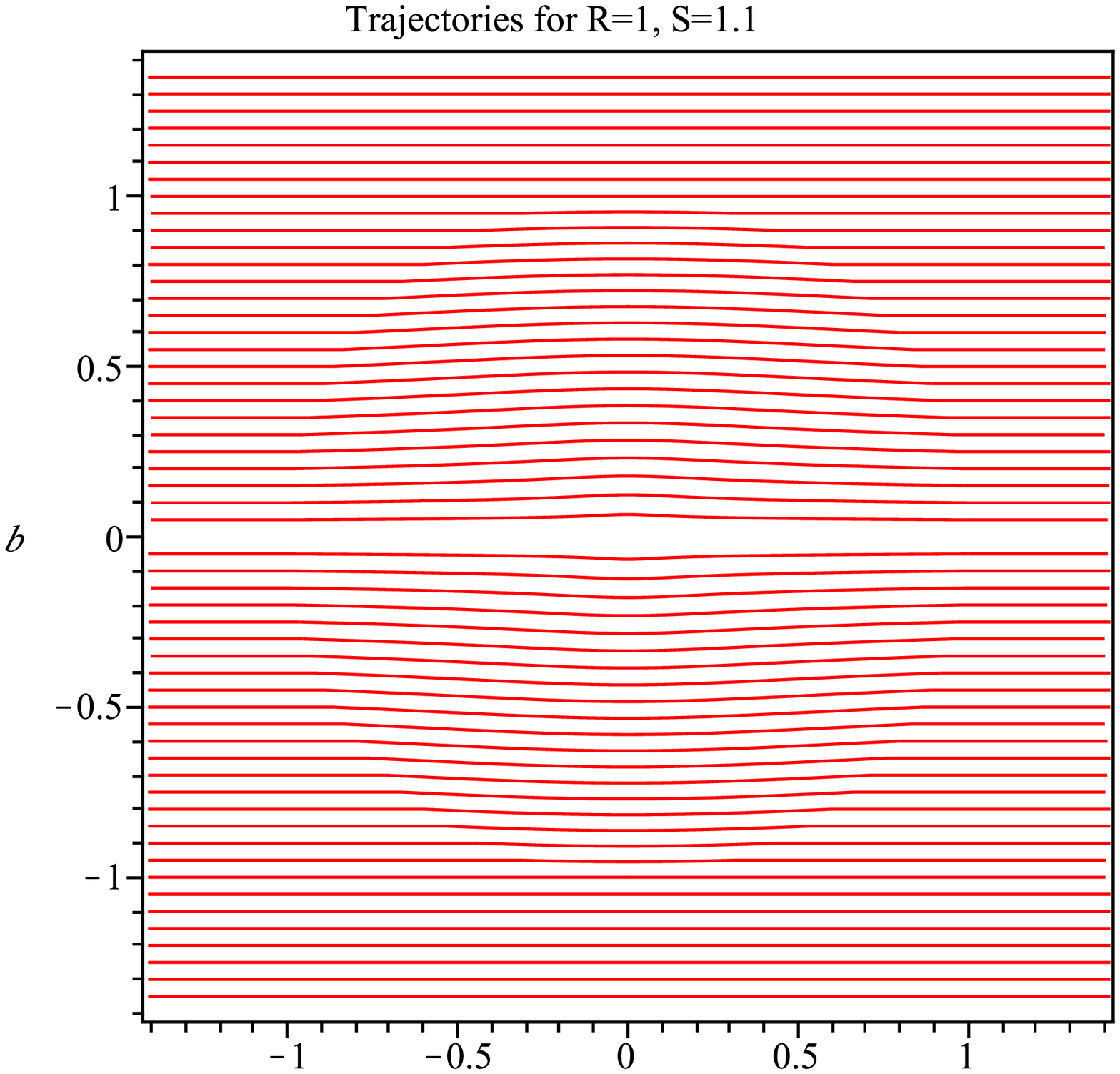} } $\; \; \;$
         \subfloat[Weak effective cloaking $S=1.5$ ]{\label{fig:RVclass} \includegraphics[trim=11mm 2mm 2mm 10mm,clip,width=0.21\textwidth]{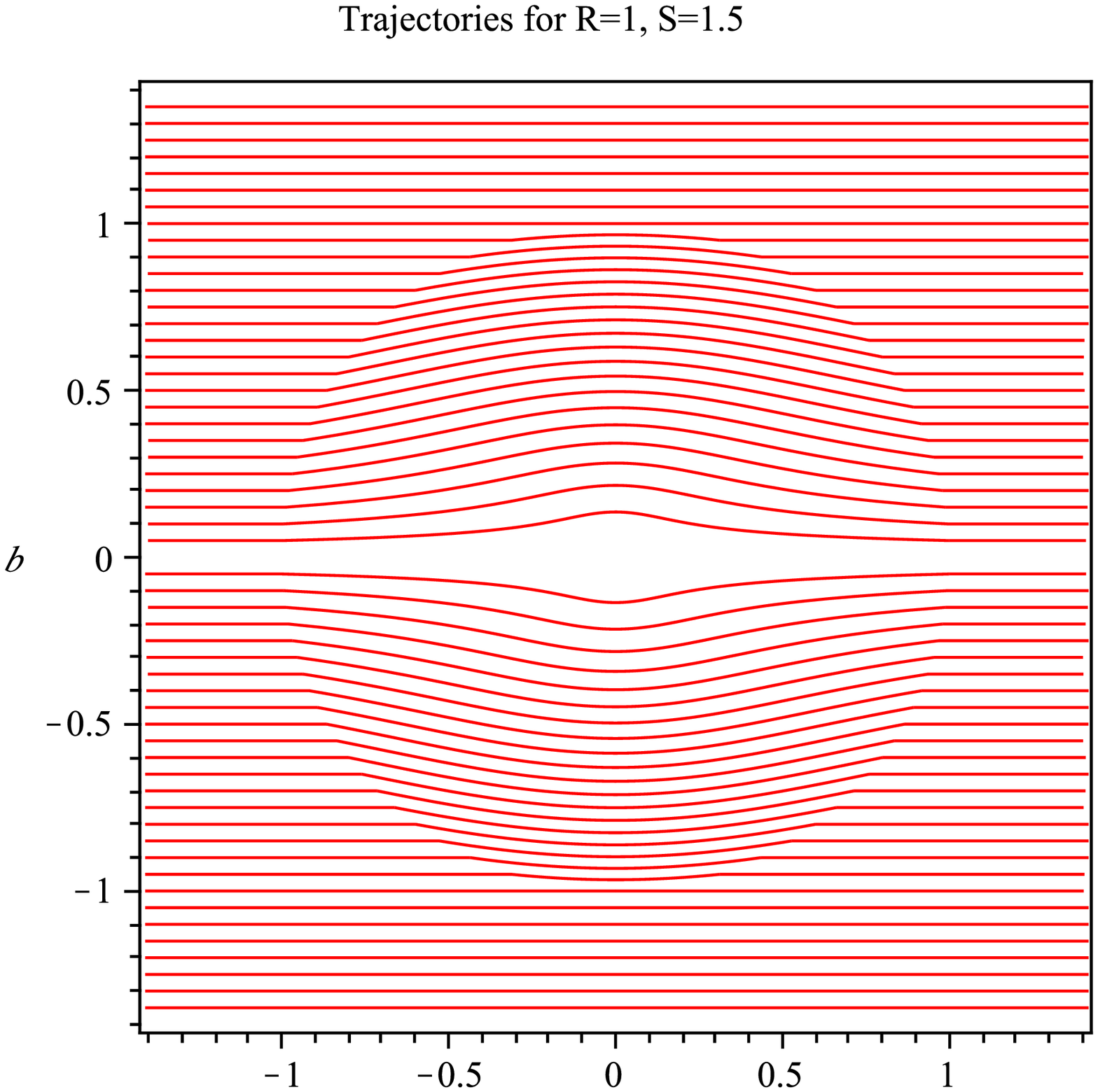} }
  
         \subfloat[Strong effective cloaking $S=3$]{\label{fig:classscalarpenrose} \includegraphics[trim=16mm 2mm 3mm 8mm,clip,width=0.22\textwidth]{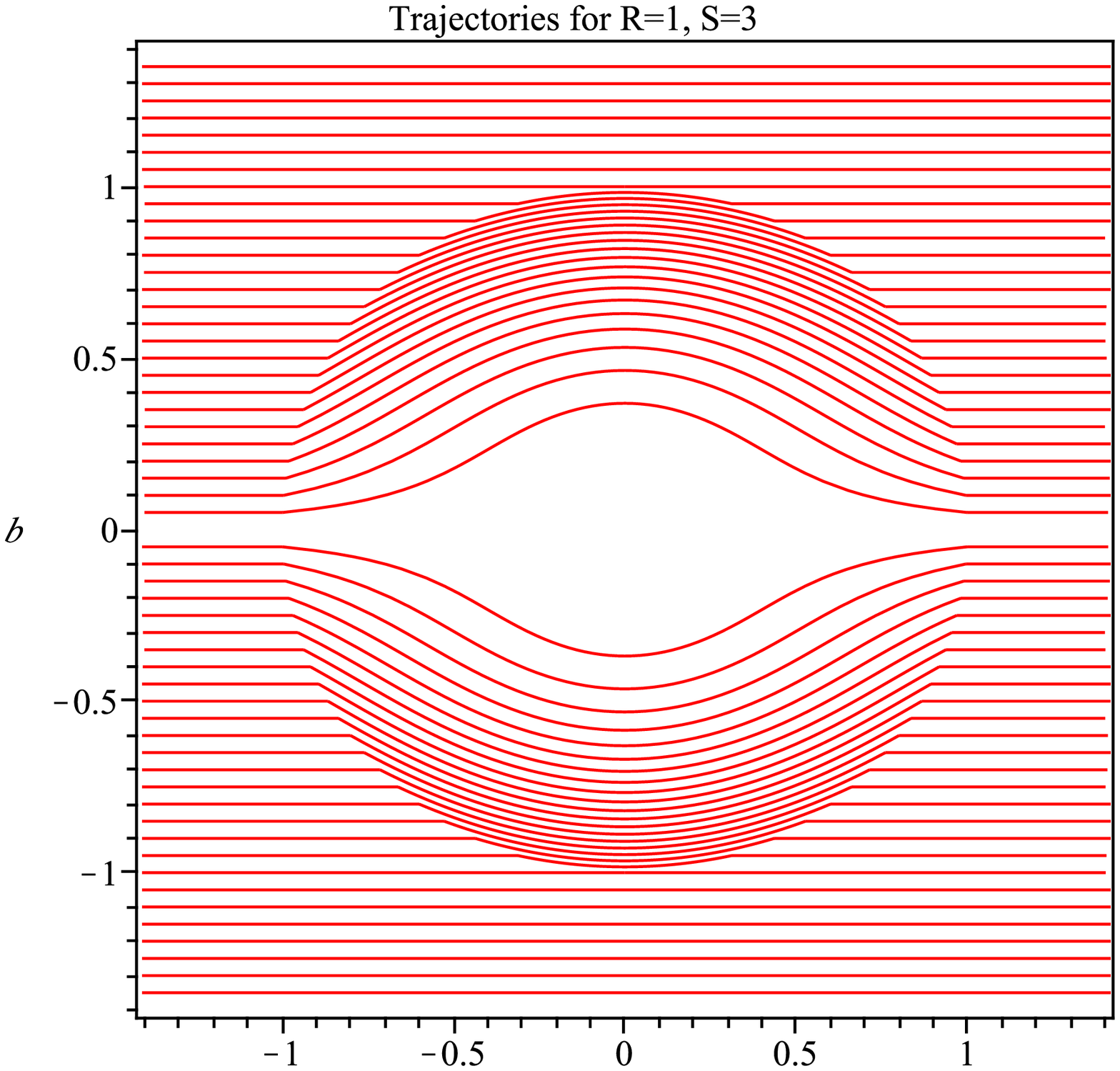} }
         \subfloat[Very strong effective cloaking $S=10$]{\label{fig:classscalarpenrose} \includegraphics[trim=10mm 5mm 1mm 9mm,clip,width=0.22\textwidth]{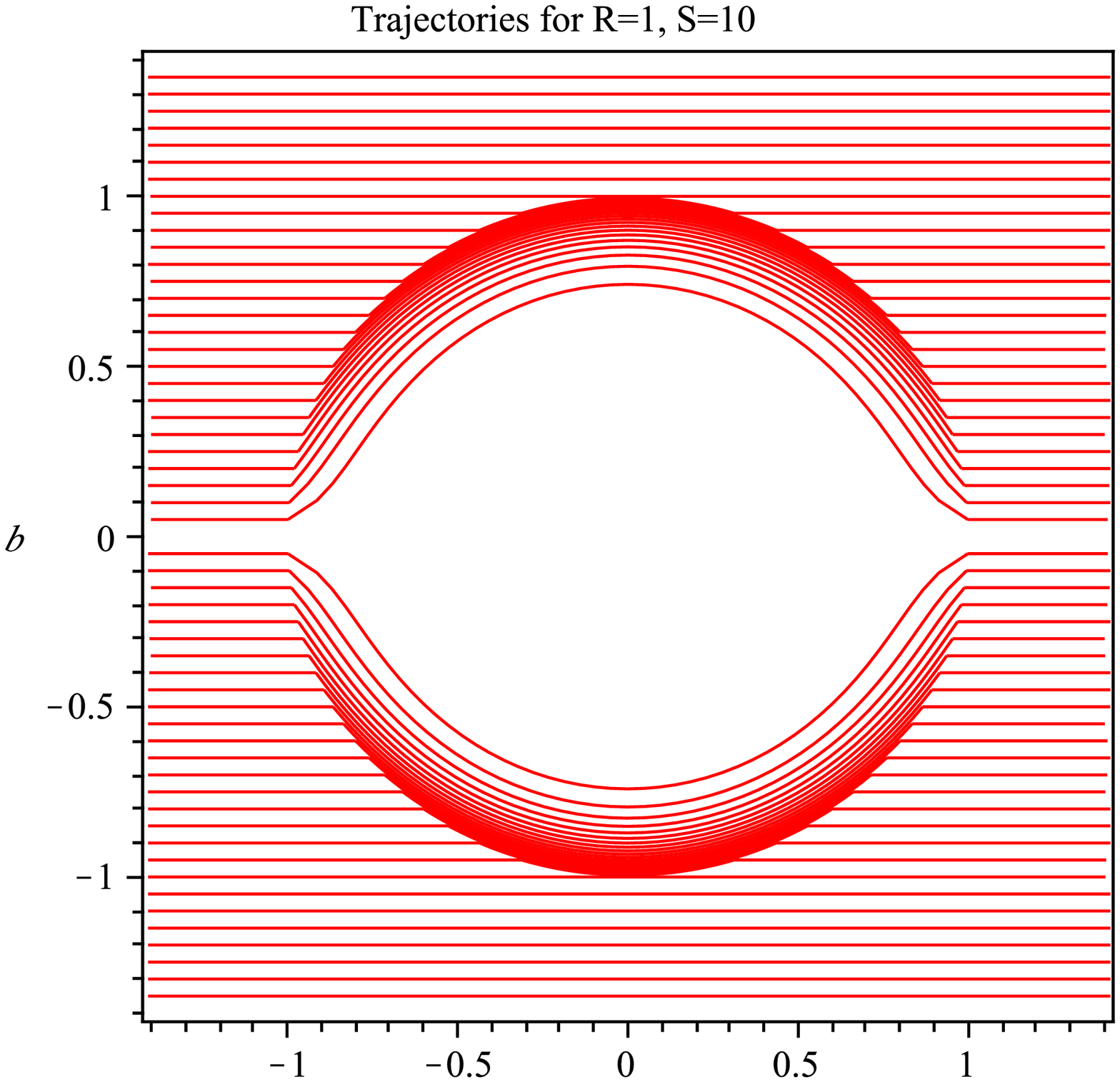} }
          \caption{Spatial trajectories of null geodesics moving across $R=1$ geometries with a variety of $S$ parameters.\label{fig:Trajectoreis} }
\end{figure}
To demonstrate the effective cloaking feature of these geometries, let us consider
the trajectories of a congruence of parallel null geodesics incident upon the
junction hypersurface from the exterior, where individual geodesics are distinguished by the \emph{impact parameter} $b$:
\begin{equation}
t=\lambda\,,\, r=\sqrt{b^{2}+\lambda^{2}}\,,\, sin\theta=\frac{\lambda}{\sqrt{b^{2}+\lambda^{2}}}\;,\;\phi=0\;,\label{eq:congruenceEXT}
\end{equation}
where $\lambda$ is the affine time. Due to the spherical symmetry of the system, the trajectory of an arbitrary null
geodesic can be described as a rotated member of this congruence. 

The geodesics in this congruence have tangent
vectors:
\begin{equation}
\xi_{OUT}^{a}=[1,\,\frac{\lambda}{\sqrt{b^{2}+\lambda^{2}}},\:\frac{b}{b^{2}+\lambda^{2}},\,0]\;.\label{eq:vecout}
\end{equation}

Since all of these geodesics lie along a great circle, we need only concern ourselves with the interior null geodesics whose
tangent vectors can be written:
\begin{equation} \begin{split}
\xi_{IN}^{a}=& [a_{1}(\frac{\tilde{r}}{SR})^{\beta},\,\pm\sqrt{a_{1}^{2}(\frac{\tilde{r}}{SR})^{\beta}-a_{2}^{2}\frac{S^{2}}{\tilde{r}^{2}}},\, a_{2}\frac{S^{2}}{\tilde{r}^{2}}\:,0]\;,
\\ \beta\equiv & -2+2S
\label{eq:vecin} \end{split}
\end{equation}
where $a_{1}$ and $a_{2}$ are constants of motion. 

We extend the exterior congruence into the interior geometry by
matching the projections of the tangent vectors of Eq.(\ref{eq:vecin})
and Eq.(\ref{eq:vecout}) on the junction hypersurface. The projected
tangent vectors on the surface $r=R$ and $\tilde{r}=SR$ are respectively:
\begin{equation}\begin{split}
P_{b}\equiv h_{ab}\xi_{OUT}^{a}=[-1,\,0,\, b,\,0]\;,\\
P_{b}\equiv h_{ab}\xi_{IN}^{a}=[-a_{1}\,0,\, a_{2},\,0]\;,
\end{split}\end{equation}
where $h_{ab}=g_{ab}-r_{a}r_{b}$, and $r_{a}$ denotes the unit normal
to the hypersurface. Thus, the constants of motion for the extended
geodesics are: $a_{1}=1$, $a_{2}=b$. The geodesics in the parallel
congruence Eq. (\ref{eq:congruenceEXT}), as they move through the
interior geometry, will therefore have tangent 4-vectors:
\[
\xi_{IN}^{a}=[(\frac{\tilde{r}}{SR})^{-2+2S},\,\sqrt{(\frac{\tilde{r}}{SR})^{-2+2S}-b^{2}\frac{S^{2}}{\tilde{r}^{2}}},\, b\frac{S^{2}}{\tilde{r}^{2}},\,0]\;.
\]

Thus, the spatial trajectory of the geodesics will have tangents satisfying:
\begin{equation}
\frac{\partial\tilde{r}(\theta)}{\partial\theta}=\frac{1}{b}\frac{\tilde{r}^{2}}{S^{2}}\sqrt{(\frac{\tilde{r}}{SR})^{-2+2S}-b^{2}\frac{S^{2}}{\tilde{r}^{2}}}\;.\label{eq:TrajAngle}
\end{equation}
Note also that the impact parameter $b$ can be re-written in terms
an angle of incidence $\theta_{i}$, describing the angle at which
an external geodesic meets the junction hypersurface: $b\equiv R\cos\theta_{i}$. 

The differential equation Eq.(\ref{eq:TrajAngle}) can be solved explicitly
through the transformation: $\tilde{r}=RS\left(\frac{1}{F(\theta)}\right)^{\frac{1}{S}}$,
where $F(\theta)$ must satisfy:
\[
\frac{\partial F}{\partial\theta}=-\frac{1}{b}R\sqrt{1-\frac{b^{2}}{R^{2}}F^{2}(\theta)}\;.
\]
The unique solution to this differential equation is: 
\[
F=\frac{R}{b}\cos\theta\:,
\]
and thus the trajectory through the interior must satisfy:
\begin{equation}
\tilde{r}(\theta)=RS\left(\frac{b}{R\cos\theta}\right)^{\frac{1}{S}}=RS\left(\frac{\cos\theta_{i}}{\cos\theta}\right)^{\frac{1}{S}}\,.\label{eq:Trajectories}
\end{equation}

Note that all geodesics in this congruence have a radial turning point halfway across the interior (at $\theta=0$): $\frac{\partial}{\partial\theta}\tilde{r}(\theta)\vert_{\theta=0}=0$.
Thus, the trajectories of the geodesics in our congruence are symmetric in reflections along the line $\theta=0$, and the null congruence will emerge from the interior geometry undeformed.  

The family of cloaking geometries have two free  parameters: the radius of the junction surface
$r=R$, and the value of the parameter $S>0$. The larger the value
of $S$ is, the larger the effect of geodesics 
splaying away from the center (see Fig. (\ref{fig:Trajectoreis})).

Since the image of the star field behind an astronomical
object with this spacetime geometry will not be distorted as the photons pass through, we refer to this geometry as an effective cloaking geometry. 

\section{Properties of the cloaking geometry}

\subsection{Reducing the shadow of an object}

\begin{figure}
\centering \protect\includegraphics[trim=53mm 55mm 43mm 53mm,clip,width=0.45\textwidth]{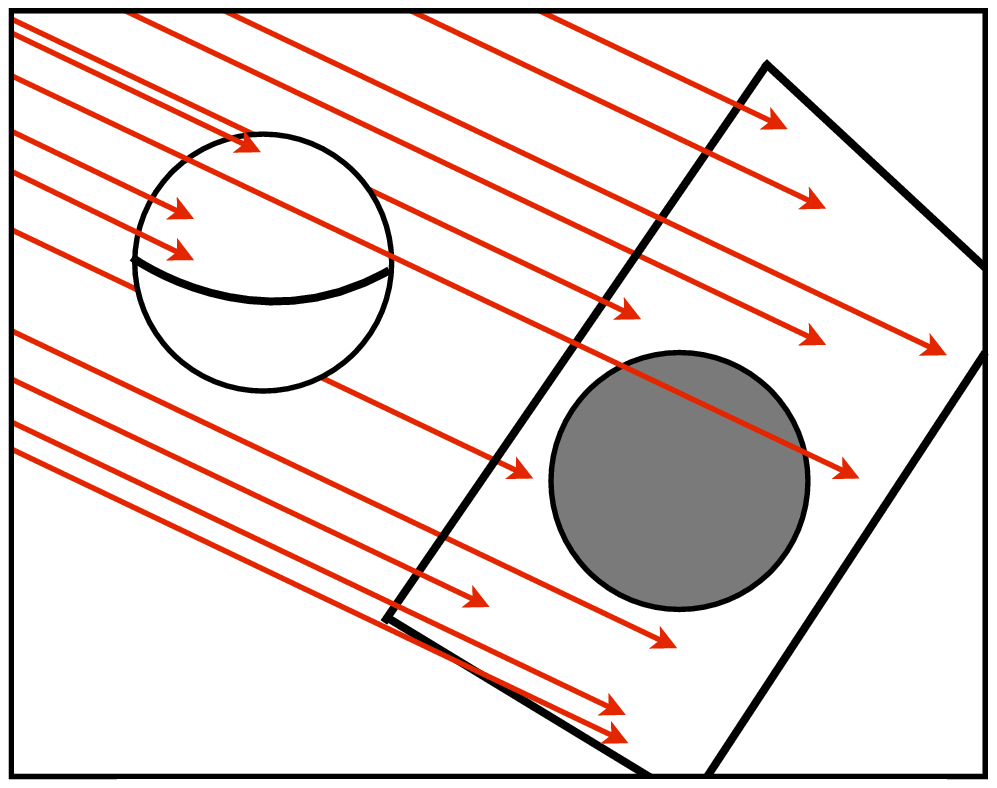}
\caption{We describe the size of an object with respect to a congruence of geodesics. The object eclipses some of the curves, resulting in a shadow whose area depends on the size of the object, its orientation, and the congruence itself. \label{fig:rain}}
\end{figure}

When we optically gauge the size of an object, we usually do so by
looking at its projected area on hypersurfaces normal to a set of
null curves which span the space between the observer and the object.
In other words, we look at the object's shadow with respect to a set
of (preferably parallel) null geodesics (see Fig.\eqref{fig:rain}). Given enough information, we  deduce the object's volume based on the area of this shadow. 

For example, imagine that we have been given a metallic sphere, of radius $r_{sphere}$,
and a congruence of parallel
null geodesics (called the \emph{background image}) which are partially eclipsed by the sphere. For all orientations, the
sphere will cast a shadow of area $A_{sphere}=\pi r_{sphere}^{2}$ in
the background image. 

What would this area be if we were to place our metallic sphere at the center of
a cloaking geometry? We can determine the area of the shadow, as
seen from the outside, by looking at the range of impact parameters $b$
for null geodesics which will be eclipsed by the ball.

In the interior coordinates, our ball will have a radius $\tilde{r}_{sphere}=\frac{r_{sphere}}{S}$.
Eq.\eqref{eq:Trajectories} dictates the trajectory of the geodesics
in our background image congruence. These geodesics' minimum radius are
at angle $\theta=0$. Thus, given $R$ and $S$, the geodesics whose impact parameter
$b$ satisfies:
\[
r_{sphere}\geq R\left(\frac{b}{R}\right)^{\frac{1}{S}}
\]
will encounter our spherical obstruction (where we assume $r_{sphere}<R$).
Thus, the edge of the shadow will have impact parameter $b=R\left(\frac{r_{sphere}}{R}\right)^{S}$,
and the area of the shadow will be:
\[
A_{effective}=\pi R^{2} \left(\frac{r_{sphere}}{R}\right)^{2S}\;.
\]
The larger the value of the parameter $S$, the smaller the shadow
of the sphere will be. 

Thus,  by tuning the value of $S$, we can use the cloaking geometry
to make an object appear arbitrarily small from the outside.

\subsection{Stress Energy Source}

The stress energy tensor of this geometry has two components:
the stress-energy generating the interior geometry, and the stress-energy
shell which lies on the junction surface:
\[T_{ab}=T^{(IN)}_{ab} +T^{\Sigma}_{ab}\,.\]

 From the Einstein equation, stress energy tensor
of the interior geometry is:
\[
T_{ab}^{(IN)}=\frac{(S-1)^2}{8\pi}\left[\begin{array}{cccc}
\frac{(S+1)(SR)^{2S-2}}{\tilde{r}^{2S}(S-1)} & 0 & 0 & 0\\
0 & \frac{(S+3)}{\tilde{r}^{2}(1-S)} & 0 & 0\\
0 & 0 & \frac{1}{S^{2}} & 0\\
0 & 0 & 0 & sin^{2}\theta\frac{1}{S^{2}}
\end{array}\right].
\]
The fact that null geodesics splay away from the center (in the interior geometry)
indicates that the null convergence condition is not satisfied,
and thus the null energy condition is violated. We can confirm
this explicitly using geodesics of the form Eq. (\ref{eq:vecin}):
\begin{equation} \begin{split}
T_{ab}^{(IN)}\xi_{IN}^{a}\xi_{IN}^{b}= \frac{2(S-1)}{8\pi}\frac{1}{\tilde{r}^{4}}\left( \right. & a_{2}^{2}(S+1)S^{2}  \\  & -(\frac{1}{SR})^{-2+2S}a_{1}^{2} \left. \tilde{r}^{2S}\right)\,.
\end{split} \end{equation}
 since parameters $a_{1}$ and $a_{2}$ are free, we can choose $a_{1}=1$,
$a_{2}=0$, demonstrating that the violation. 

The Ricci scalar in the interior is explicitly:
\[
R=\frac{6(S-1)}{\tilde{r}^{2}}\;,
\]
demonstrating the existence of a central curvature singularity.

The nonzero components of the second fundamental form on the interior
of the junction sphere are:
\[
K_{tt}^{(IN)}=\frac{S-1}{RS}\,,\, K_{\theta\theta}^{(IN)}=\frac{R}{S}\,,\, K_{\phi\phi}^{(IN)}=\frac{Rsin^{2}(\theta)}{S}\,,
\]
 and from the exterior:
\[
K_{tt}^{(OUT)}=0\,,\, K_{\theta\theta}^{(OUT)}=R\,,\, K_{\phi\phi}^{(OUT)}=Rsin^{2}(\theta)\,.
\]
 The Israel junction conditions dictate that the discontinuity in
the second fundamental form describes a shell of matter located at the junction. A 3-tensor
on the hypersurface is defined:
\begin{equation}\begin{split}
S_{ij}\equiv & \frac{1}{8\pi}([K_{ij}]-[K]h_{ij}) \\ &=\frac{1}{8\pi}(\frac{S-1}{S})\left[\begin{array}{ccc}
\frac{2}{R} & 0 & 0\\
0 & -2R & 0\\
0 & 0 & -2Rsin^{2}\theta
\end{array}\right]\,.
\end{split}\end{equation}
 This is used to write the energy density of the shell of matter
lying on the junction hypersurface:
\begin{equation}\begin{split}
T_{ab}^{\Sigma}\equiv & \delta(\ell)S_{ij}e_{a}^{i}e_{b}^{j} \\ =  & -2R\frac{\delta(\ell)}{8\pi}(\frac{S-1}{S})\left[\begin{array}{cccc}
-\frac{1}{R^{2}} & 0 & 0 & 0\\
0 & 0 & 0 & 0\\
0 & 0 & 1 & 0\\
0 & 0 & 0 & sin^{2}\theta
\end{array}\right]\,,
\end{split}\end{equation}
where $\ell$ is the length along a congruence normal to the junction, and $\ell\equiv0$ at the junction.

\subsection{Redshifting}

Let us consider a family of stationary timelike observers sitting at constant
radii with 4-vectors:
\[
T^{a}=[(\frac{\tilde{r}}{SR})^{S-1},\,0,\,0,\,0]\,,
\]
and look at the redshifts of null geodesics passing between them.
In the flat exterior, let us denote the 4-momentum of a photon $P^{a}$ and the energy of a photon 
measured by a stationary observer to be $\gamma_{i}$:
\[
P_{OUT}^{a}T^{b}g_{ab}=-\gamma_{i}\,.
\]
As the photon travels into the cloaking geometry, the stationary
observer measures the photon energy to be:
\[
\gamma_{f}=\gamma_{i}(\frac{\tilde{r}}{SR})^{-1+S}\;.
\]
Thus, the photon is redshifted as it moves towards the center, and
blueshifted as it moves outwards.

\section{Conclusion}

We have presented a two parameter family of spacetimes which demonstrate the effective cloaking of objects placed within them as a
result of the gravitational lensing. The analytic
trajectory of the null geodesics moving through the cloaking geometry
is known, and it is shown that initially parallel geodesics entering
the cloaking geometry will splay away from the center,  re-converge,
and then reemerge in their original, parallel
configuration. 

The description of this geometry as \emph{effectively cloaking} derives
from the way in which an object placed at its center will eclipse fewer null
geodesics than it would in flat spacetime. This is
due to the splaying of the null geodesics away from the center and around the object. Thus, we
use the geometry to make an object appear arbitrarily small from the outside.

Several attributes of this geometry make it arguably physically unrealizable.
Firstly, the matter used to construct it must violate the null (and thus the weak and dominant) energy
condition. Secondly, this geometry requires an infinitesimally narrow
shell of stress-energy to transition between interior and exterior geometries, and it is unclear what effect allowing a transition of finite width will have on the cloaking properties. The requirement for exotic energy is, however, the the same shortcoming 
found in traversable wormholes and warp drive spacetimes.

\section*{Acknowledgements}

This work was supported by the Natural Science and Engineering Research
Council of Canada. I would like to thank Dr. Viqar Husain for introducing
me to the topic and for his invaluable insights. 

\bibliographystyle{plain}
\bibliography{cloak2}

\end{document}